\documentclass[aps,prb,twocolumn,floatfix]{revtex4-2}
\usepackage[colorlinks=true,citecolor=blue]{hyperref}
\usepackage{graphicx}% Include figure files
\usepackage{dcolumn}% Align table columns on decimal point
\usepackage{bm}% bold math
\usepackage[export]{adjustbox}
\usepackage{lipsum}
\usepackage{braket}
\usepackage{eucal}
\usepackage{color}
\usepackage{bm}
\usepackage{bbold}
\usepackage{amsmath}
\usepackage{amssymb}
\usepackage{mathtools}
\usepackage{bigints}
\usepackage{float}
\usepackage{titlesec}
\usepackage{gensymb}
\usepackage{gensymb}
\usepackage{appendix}

\def\be{\begin{equation}}
\def\ee{\end{equation}}
\def\bea{\begin{eqnarray}}
\def\eea{\end{eqnarray}}
\def\v{{\bf v}}
\def\ev{{\bf E}}
\def\bv{{\bf B}}

\def\k{{\bf k}}
\def\q{{\bf q}}

\def\m{{\bf m}}
\def\j{{\bf j}}

\def\rv{{\bf r}}

\def\s{{\boldsymbol\sigma}}
\def  \bsig    {\mbox{\boldmath$\sigma$}}
\usepackage{color}
\begin{document}

\title{Topological charge and spin Hall effects due to skyrmions in canted antiferromagnets}

\author{A. N. Zarezad$^{1}$, A. Qaiumzadeh$^{2}$, J. Barnaś$^{1}$, A. Dyrdał$^{1}$}
\email[]{adyrdal@amu.edu.pl}
\affiliation{$^{1}$ Department of Mesoscopic Physics, ISQI, Faculty of Physics,
 Adam Mickiewicz University, ul. Uniwersytetu Poznańskiego 2, 61-614 Poznań, Poland\\
 $^{2}$ Center for Quantum Spintronics, Department of Physics, Norwegian University of Science and Technology, NO-7491 Trondheim, Norway
}

\begin{abstract}
The topological charge-Hall effect (TCHE) and the topological spin-Hall effect (TSHE), arising from ferromagnetic (FM) and antiferromagnetic (AFM) skyrmions, respectively; can be elucidated through the emergence of spin-dependent Berry gauge fields that affect the adiabatic flow of electrons within the skyrmion texture. TCHE is absent in systems with parity-time (PT) symmetry, such as collinear AFM systems. 
In this paper, we theoretically study TCHE and TSHE in a canted antiferromagnet within the diffusive transport regime. Spin canting or weak ferromagnetism in canted AFMs, which break the PT symmetry, may arise, e.g., from strong homogeneous Dzyaloshinskii-Moriya interactions.
Using a semiclassical Boltzmann approach, we obtain diffusion equations for the spin and charge accumulations in the presence of finite spin-flip and spin-dependent momentum relaxation times. We show that the weak ferromagnetic moment stemming from spin canting and the subsequent breaking of parity-time symmetry, results in the emergence of both finite TCHE and TSHE in AFM systems.
\end{abstract}

\date{\today}


\maketitle

%----------------------------------------------------------------------------------------
\section{Introduction}
\label{sec:intro}
%----------------------------------------------------------------------------------------

Recently, there has been great interest in various topologically nontrivial magnetic
textures, such as skyrmions. Skyrmions are stable topological solitons arising from certain classes of non-linear sigma models, as was formulated long time ago in field theory; see Refs.~\cite{samoilenka2017gauged,skyrme1994non} and references therein. The existence of skyrmions in magnetic systems was later introduced as a metastable state in isotropic ferromagnetic (FM) systems \cite{belavin1975metastable} and later as either metastable state (single skyrmions) or stable state (skyrmion crystals) in the presence of chiral spin interactions, such as the Dzyaloshinskii-Moriya interaction (DMI)~\cite{BOGDANOV1994255,PhysRevLett.87.037203,Bogdanov,Kiselev_2011, Leonov_2016,PhysRevB.96.140411}. 
Skyrmions were first experimentally discovered in a chiral magnet in 2009 \cite{science.1166767}. 

Many aspects of magnetic skyrmions, including their stability, dynamics, excitations, etc., were addressed in the recent studies~\cite{wilson2014chiral,seki2012observation,leonov2015multiply,psaroudaki2017quantum,schutte2014magnon,white2014electric}. This not only concerns single skyrmions but also skyrmion lattices, called skyrmion crystals~\cite{Yu2010,miao101063,PhysRevLett.108.017601,PhysRevB.103.064414,PhysRevB.106.104424}. Of particular interest from the fundamental point of view were the topological properties of skyrmions, which are now well understood and well described by appropriate topological parameters.  

An important practical issue is the control of skyrmion positions, including also the control of their motion. A single skyrmion can be pinned to a certain pinning center. For instance, magnetic nanodots in an overlayer covering a magnetic film with skyrmions, may serve as pinning centers, and the skyrmions become then confined in the regions below the nanodots. They may perform spiral  clockwise or anticlockwise motion in these regions, and the winding direction depends on the confining field. By reversing this field one may 
reverse the skyrmion winding trajectories,
and this may lead to the skyrmion echo~\cite{PhysRevLett.129.126101}, similar to the well known spin echo~\cite{abragam1961principles,PhysRev.98.1105}.

An interesting way of dynamical pinning (and thus also of skyrmion motion) can be realized when the magnetic material hosting skyrmions displays magnetoelectric coupling in a noncollinear phase (within skyrmions). Then, owing to the magnetoelectric coupling, a laser beam becomes a pinning center for skyrmions. Moving the laser beam effectively moves the pinned skyrmion in a fully controlled way~\cite{Wang2020}. Instead of a laser beam, one can use the electric field of surface plasmon polaritons in an attached metallic layer to create plasmonic lattice. The nodes of this lattice are pinning centers for skyrmions, attracting them to form a plasmonic-skyrmion lattice~\cite{PhysRevLett.125.227201}.    

The most important way of controlling the skyrmion motion is by an external electric field (or effectively by current). 
It is well known that
spin-polarized current flowing along the nanoribbon with skyrmions drags the skyrmions along the electric field and also deflects their trajectories towards one of the nanoribbon edges (the skyrmion Hall effect) ~\cite{Liang2015,nphys2231, Nagaosa2013, Iwasaki2013, Litzius2017, nphys3883, PhysRevB.95.094401}. This phenomenon was studied both theoretically and  experimentally, see Refs.~\cite{Litzius2017, nphys3883, PhysRevLett.100.127204, PhysRevB.78.134412, PhysRevLett.107.136804, Hans2012, Tomasello2014} for an overview.  
In turn, FM skyrmions deflect electron trajectories in the direction perpendicular to the external electric field~\cite{PhysRevLett.93.096806, PhysRevLett.102.186602, PhysRevLett.102.186601, PhysRevB.91.245115, PhysRevLett.117.027202, PhysRevB.95.064426,PhysRevB.97.134401}. This phenomenon is qualitatively similar to the anomalous Hall effect in FM metallic layers with uniform magnetization \cite{ PhysRev.95.1154, RevModPhys.82.1539, PhysRevLett.112.017205}. The origin of this skyrmion-induced topological charge Hall effect (TCHE) is the emergence of a real-space Berry curvature induced by the skyrmion textures ~\cite{ PhysRevLett.93.096806, PhysRevLett.98.246601}, while in the anomalous Hall effect, the Berry curvature emerging in the  momentum space due to the spin-orbit couplings ~\cite{ PhysRevB.64.104411, PhysRevB.64.014416, PhysRevB.68.045327}.

The situation is different in the case of antiferromagnetic (AFM) skyrmions, where there is no skyrmion Hall effect \cite{10.1063/1.4967006, PhysRevLett.116.147203, Zhang2016_1, Zhang2016, PhysRevB.99.054423, TRETIAKOV2021333, GOBEL20211, Amin2023}. This happens as the net perpendicular driving force exerted on a skyrmion vanishes in the case of parity-time (PT) symmetric AFM system. A similar situation also occurs in the case of skyrmions in two FM layers coupled  antiferromagnetically by the interlayer exchange interaction, the so-called synthetic AFM systems \cite{PhysRevB.96.060406,Legrand2020}. The lack of deflection of the AFM skyrmions is one of the advantages of AFM systems over the FM ones in the context of practical applications in spintronic devices.  
It was theoretically shown that AFM skyrmions may also create a real-space Berry curvature \cite{PhysRevB.86.245118}, leading to topological spin Hall effect (TSHE) in AFM systems~\cite{RRL1700007, PhysRevB.96.060406, PhysRevLett.121.097204, nakazawa2023topological}. These papers also confirmed the absence of TCHE in these systems. The considerations were based on simple square and hexagonal AFM lattices with PT symmetry.

In our recent work \cite{ZAREZAD2024171599}, we have revisited both TCHE and TSHE in a collinear square-lattice AFM system. We considered finite asymmetric spin-dependent scattering. Such an asymmetry may appear when the system is intentionally doped with magnetic scattering centers. Our description was based on the Boltzmann kinetic equation with an emerging magnetic field due to skyrmions included in a diffusive regime. As a result, we found not only a finite TSHE, but also a finite TCHE. However, the latter effect disappears when the asymmetry in the spin-dependent relaxation times vanishes, in agreement with earlier studies.     

\begin{figure}[t]
%	\centering
  		\includegraphics[width=.8\columnwidth]{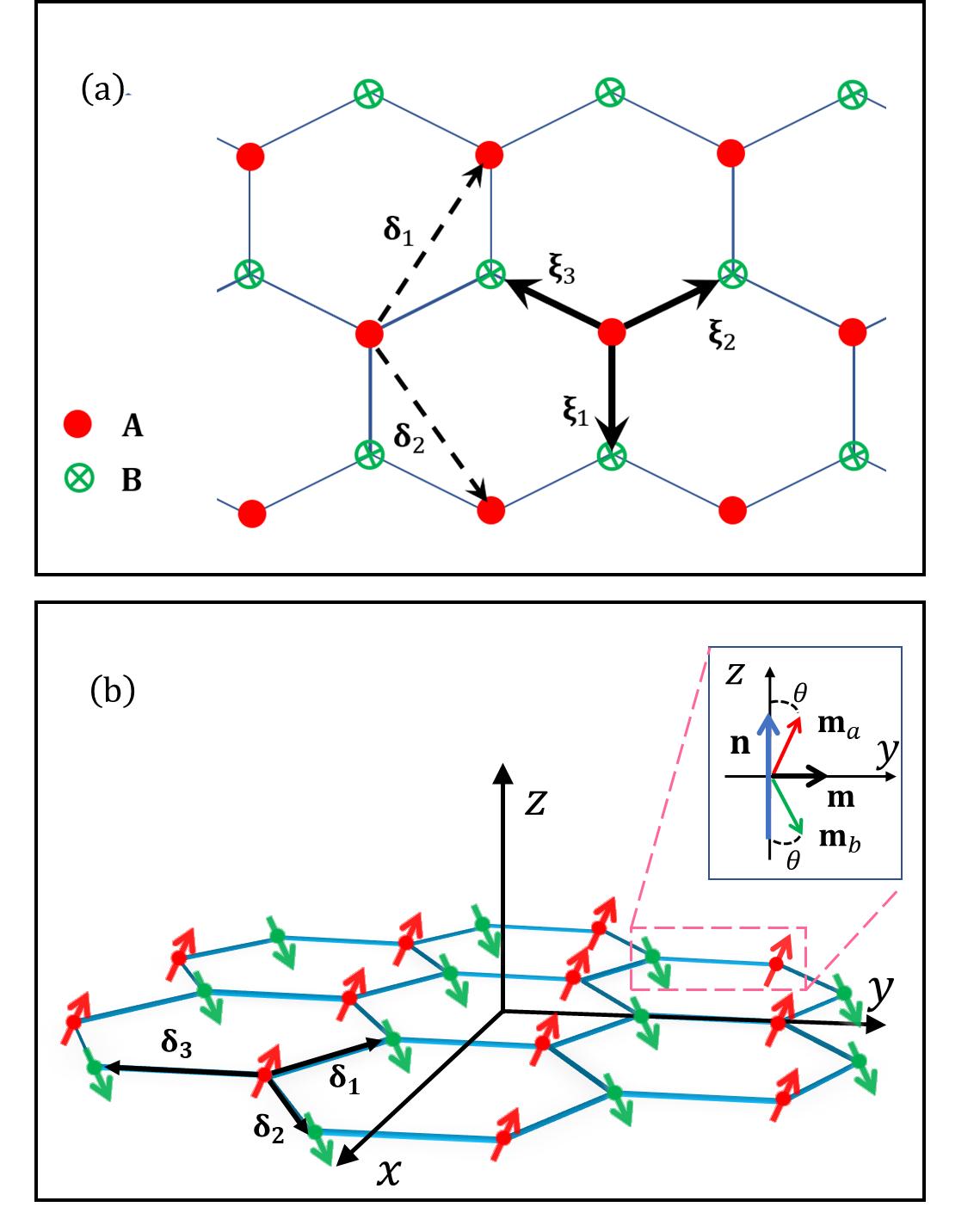}
	\caption{(a) Schematic representation of a hexagonal lattice  consisting of two sublattices, A and B, with two lattice unit vectors ${\bm\delta}_1$ and ${\bm\delta}_2$, and the three nearest-neighbor vectors,  ${\bm \xi}_i$.
     (b) Schematic configuration of the local magnetic moments, ${\bm m}_a$ and ${\bm m}_b$, in a canted AFM hexagonal lattice with a uniform canting angle $\theta$, that induces a net magnetization $m=\sin\theta$. The inset shows the net magnetization and staggered sublattice moments. }
	\label{fig:Honeycomb-lattice}
\end{figure}

Recent studies have shown that chiral anomalous Hall effects may appear in canted AFM systems \cite{Kipp2021, Qiang_2022}. Furthermore, experimental results obtained on ${\mathrm{Ca_{1-x}Ce_{x}MnO_3}}$ have confirmed the existence of the Hall effect in the canted AFM systems \cite{ Vistoli2019}. The number of articles that examine this issue is very limited.  One such article is Ref. \cite{ PhysRevB.101.174432}, where the authors considered a square lattice and treated the uniform magnetic moment induced by a canting of local spins as a perturbation. Using Kubo formalism, they derived an analytical expression for the topological charge Hall conductivity. 

In this paper, we investigate skyrmion-induced TSHE and TCHE in a canted hexagonal AFM system, with the weak ferromagnetic moment due to canting of the AFM sublattice magnetic moments. We assume that the magnetization inside the skyrmion is non-uniform and perpendicular to the N{\'e}el vector, following the same profile as the N{\'e}el vector. This extends recent studies on skyrmions in compensated AFM systems to weak ferromagnets~\cite{DZYALOSHINSKY1958241, PhysRev.120.91, 10.1063/1.4865565, doi:10.1142/2518}. Unlike Ref. \cite{ PhysRevB.101.174432}, we include the ferromagnetic moment nonperturbatively, and use semiclassical Boltzmann kinetic approach with both spin-dependent relaxation times and spin-flip scatterings. From the Boltzmann equation, we find analytical expressions for both spin and charge currents along the orientation normal to the driving current as well as spin accumulation at the edges in the presence of a single AFM skyrmion. 
We show that owing to a net FM moment in the system, the TCHE appears even in the absence of spin-asymmetric scattering processes.

The rest of the paper is structured as follows: In Sec. \ref{sec:model}, we introduce our model Hamiltonian for a canted AFM system on a hexagonal lattice. In Sect. \ref{sec:em_field}, we compute the emergent magnetic fields, induced by skyrmions in the canted AFM system. In the Sec. \ref{sec:Boltzmann}, we develop  Boltzmann formalism in the presence of  emergent magnetic fields to compute TSHE, TCHE, and spin accumulations. We summarize and conclude our results in Sec. \ref{sec:Summary}.

%----------------------------------------------------------------------------------------
\section{Model Hamiltonian}
\label{sec:model}
%----------------------------------------------------------------------------------------

We consider a metallic canted AFM system consisting of two sublattices A and B, with the corresponding magnetic moment unit vectors $\mathbf{m}_a$ and $\mathbf{m}_b$, respectively, on a hexagonal lattice, see Fig. \ref{fig:Honeycomb-lattice}(a). The total Hamiltonian of the system, $\mathcal{H}=\mathcal{H}_{0}+\mathcal{H}_{\rm sd}$, consists of the electronic Hamiltonian $\mathcal{H}_{0}$ and an interacting term $\mathcal{H}_{\rm sd}$. The latter describes interaction between itinerant electrons and localized magnetic moments. These two Hamiltonians can be modelled by the following tight-binding ones;
\begin{equation}
%\begin{aligned}
\label{main_Hamiltonian_a}
\mathcal{H}_0 = -t\sum_{\rv \in  A}\sum_{i=1}^{z}\sum_{\sigma}\Big[a_{\sigma}^{\dagger}(\rv)b_{\sigma}(\rv+\bm{\xi}_i)+b^{\dagger}_{\sigma}(\rv+\bm{\xi}_i)a_{\sigma}(\rv)\Big],
\end{equation}
\begin{eqnarray}
\mathcal{H}_{\rm sd}=-
J\sum_{\sigma \sigma'}\Big[\sum_{\rv \in  {A}}\m_a(\rv)\cdot \s_{\sigma \sigma'} a^{\dagger}_{\sigma}(\rv)a_{\sigma'}(\rv) \nonumber\\
+\sum_{\rv \in  {B}}\m_b(\rv)\cdot \s_{\sigma \sigma'} b^{\dagger}_{\sigma}(\rv)b_{\sigma'}(\rv)\Big],
%\end{aligned}
\label{main_Hamiltonian_b}
\end{eqnarray}
where $a$ and $b$ ($a^{\dagger}$ and $b^{\dagger}$) are the fermionic annihilation (creation) operators of electrons belonging to two AFM sublattices A and B, respectively; $t$ and $J$ are the hopping parameter and sd exchange integral, respectively; $\bm{\xi}_i$ denotes the nearest-neighbor unit vector, and $z$ is the coordination number.
The three nearest-neighbor vectors in a hexagonal lattice, defined in Fig. \ref{fig:Honeycomb-lattice}(a), are given by,
$\bm{\xi}_1=(0,-1)a_0$, $\bm{\xi}_2=\left(\sqrt{3}/2,1/2\right)a_0$, and $\bm{\xi}_3=\left(-\sqrt{3}/2,1/2\right)a_0$,
with $a_0$ being the lattice constant.

In a general case, the sublattice magnetizations in Eq. (\ref{main_Hamiltonian_b}) are nonuniform, e.g., due to skyrmion textures. Therefore, we find first the electronic spectrum for uniform sublattice magnetizations (in the absence of skyrmions), while the general case will be considered in the subsequent section. Accordingly, we write the corresponding Hamiltonian in the momentum space as,
\begin{equation}
 \mathcal{H}_{0}=\sum_{\k \sigma}\Big[\gamma_\k a_{\sigma}^{\dagger}(\k)b_{\sigma}(\k)+c.c.\Big],
\label{main_Hamiltonian_momentum_space_a}
\end{equation}
\begin{eqnarray}
\mathcal{H}_{\rm sd}=-J\sum_{\k\sigma \sigma'}\Big\{\m_a\cdot \s_{\sigma \sigma'} a^{\dagger}_{\sigma}(\k)a_{\sigma'}(\k) \nonumber \\
+\m_b\cdot \s_{\sigma \sigma'} b^{\dagger}_{\sigma}(\k)b_{\sigma'}(\k)\Big\},
\label{main_Hamiltonian_momentum_space_b}
\end{eqnarray}
where $\gamma_\k=-t\sum_{i=1}^{z}e^{i \k \cdot \bm{\xi}_i}$ is the lattice structure factor. For a hexagonal lattice we find,
\begin{eqnarray}
|\gamma_{\k}|^2 =& t^2\Big[ 3+4\cos\left(\frac{\sqrt{3} }{2}a_{0}k_{x} \right)\cos\left(\frac{3}{2} a_{0}k_{y} \right)\Bigg. \nonumber \nonumber \\& 
+\Bigg.2\cos\left(\sqrt{3}a_0k_x\right)\Big].
\end{eqnarray}

In AFM systems, it is more convenient to introduce magnetization $\mathbf{m}=(\mathbf{m}_a+\mathbf{m}_b)/2$ and 
N{\'e}el  $\mathbf{n}=(\mathbf{m}_a-\mathbf{m}_b)/2$ vectors, where $\mathbf{m}\cdot \mathbf{n}=0$ and $\mathbf{m}^2+\mathbf{n}^2=1$.
They can be expressed in terms of the canting angle $\theta$,
\begin{align*}
\mathbf{n}=(0,0,\cos\theta), \quad \mathbf{m}=(0,\sin\theta,0),
\end{align*}
for the geometry defined in Fig.\ref{fig:Honeycomb-lattice}(b).

\begin{figure}[t]
\includegraphics[width=.79\columnwidth]{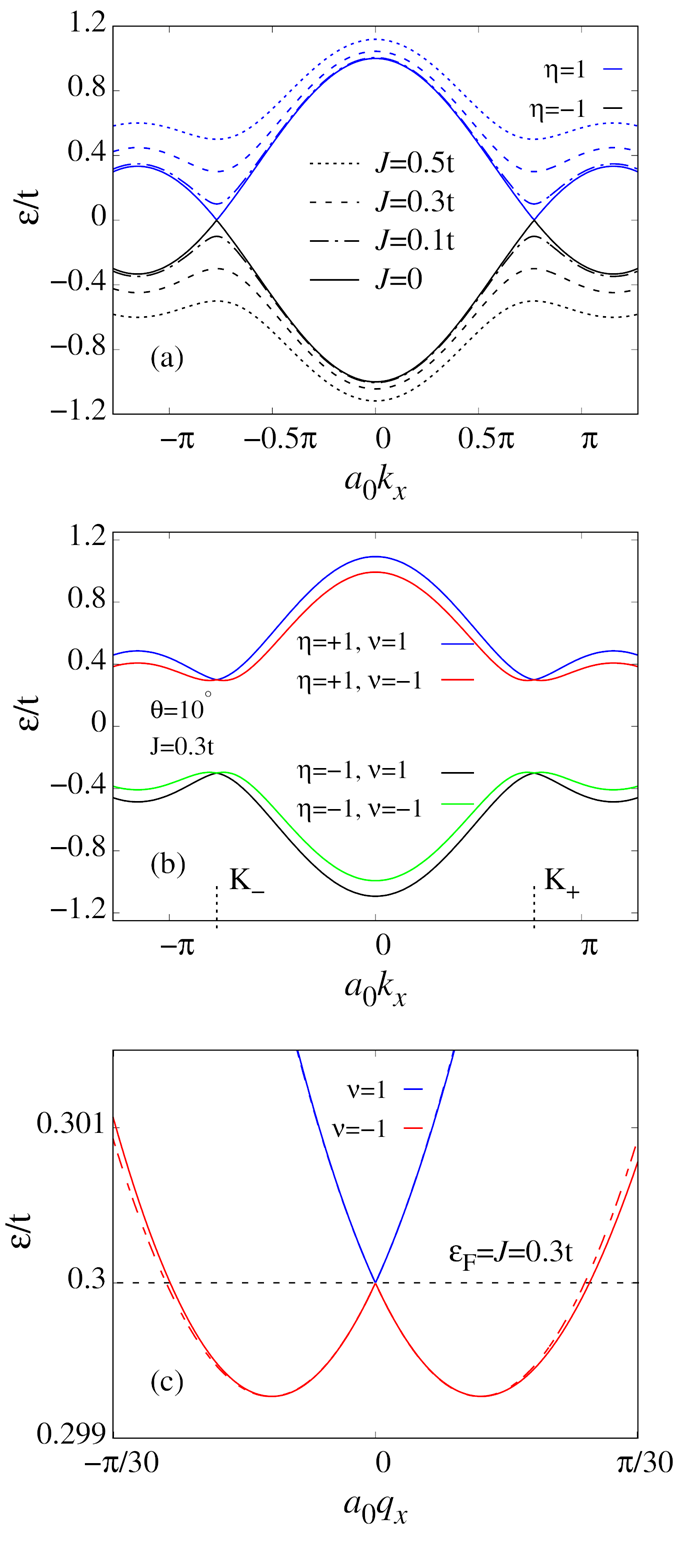}
\caption{(a) Electronic band structure of a collinear AFM system ($\theta =0$) with $J/t=0.1$, $J/t=0.3$, and $J/t=0.5$, and the nonmagnetic graphene limit $J=0$. In these cases, the conduction ($\eta =1$) and valence ($\eta =-1$) bands are spin degenerate. (b) Electronic band structure of a canted AFM system for $J/t=0.3$. Now,  the conduction and valence bands are splitted into two spin-subbands due to a small magnetic moment, $m=\sin\theta$. (c) Electronic spectrum of the two spin-subbands in the conduction band of a canted AFM system around the $K_+$ point. The solid and dashed lines correspond to the tight binding, Eq. (\ref{full_bands-canted}), and linearized, Eq. (\ref{linearizedE}), eigenenergies, respectively. 
}
 \label{fig:espectrum_AFM}
\end{figure}

In the collinear limit, $\m_a=-\m_b$ and the net magnetization and canting angle are zero, $|\mathbf{m}|=0$ and $\theta=0$. The corresponding electronic spectrum consists of spin-degenerate conduction, $\eta =+1$, and valence, $\eta =-1$, bands,
\begin{equation}
\varepsilon_{\eta}(\k)=\eta\sqrt{J^2+|\gamma_{\k}|^2}.
\label{full_bands}
\end{equation} 
The electronic dispersion in the absence of $J$ is similar to graphene, with gapless Dirac-like spectra around $K_\pm$ points, see Fig. \ref{fig:espectrum_AFM}(a) for $J=0$. The sd exchange energy $J$ opens an electronic band gap of $2J$ at these Dirac points, see Fig. \ref{fig:espectrum_AFM}(a) for $J>0$.

However, in the canted AFM case, $\m_a\nparallel\m_b$, there is a net equilibrium magnetization in the system, that lifts the spin degeneracy of both conduction and valence bands, 
\begin{equation}
\varepsilon_{\eta,\nu}(\k)=\eta\sqrt{J^2+|\gamma_{\k}|^2+2\nu Jm|\gamma_{\k}| },
\label{full_bands-canted}
\end{equation} 
where $m=|\mathbf{m}|$, and  $\nu =\pm 1$ denotes the two spin states corresponding to the quantization axis along the vector $\mathbf m$. The above dispersion equation shows that the net magnetization in the canted AFM system lifts the spin degeneracy of conduction and valence bands by shifting spin subbands in opposite directions along the momentum axis, see Fig. \ref{fig:espectrum_AFM}(b,c), resembling a Rashba-type splitting of spin subbands.

In the following, we are interested in the low-energy dynamics of electrons around Dirac points at $\mathbf{K}_\pm=\pm({4\pi}/{3\sqrt{3}},0)$ symmetric points of the Brillouin zone. Expanding the total Hamiltonian around these two points, $\mathbf{k}=\mathbf{K}_\pm +\mathbf{q}$, we find a block-diagonal Hamiltonian \cite{PhysRevLett.98.266402, PhysRevLett.98.186809,RevModPhys.81.109},
\begin{equation}
\mathcal{H}=\begin{bmatrix}
H_{_+} & 0  \\
0& H_{_-}  \\
\end{bmatrix},
\end{equation}
where $H_{\pm}$ describe the effective Hamiltonians around the $K_{\pm}$ points, 
\begin{equation}
H_{\pm} = \pm\hbar v_{0}\big(q_x \hat{\tau}_x\mp q_y \hat{\tau}_y\big)\otimes\hat{\mathbb{1}}-J\big(\hat{\tau}_z\otimes\mathbf{n}+\hat{\mathbb{1}}\otimes\bf m\big)\cdot\s.
\label{Hamiltonian}
\end{equation}
Here, $v_{0}= 3t a_0 /2\hbar$ is the Fermi velocity of gapless Dirac fermions, and $\bm{\tau} (\bm{\sigma})$ is the vector of Pauli matrices for the sublattice (spin) degree of freedom. 

The linearized eigenvalues in case of collinear AFM  and the corresponding eigenvectors can be written as, 
\begin{equation}
\varepsilon_{\eta}(\q)=\eta\sqrt{\hbar^2 v_{0}^2 q^2+J^2},
\label{eigenvlue}
\end{equation}
 \begin{equation}
\ket{\psi_{\eta}^{\sigma}}=\dfrac{1}{\sqrt{2}}\Big(\sqrt{1+\sigma\eta \mathrm{P}_{\q}}\ket{A}+\eta e^{-i\phi}\sqrt{1-\sigma\eta \mathrm{P}_{\q}}\ket{B}\Big)\otimes\ket{\bf \sigma},
\label{eigenfunction}
\end{equation}
where $\phi=\arctan({q_y}/{q_x})$, $\sigma=+1 (\uparrow )$ and $\sigma=-1 ( \downarrow$) correspond to the spin-up and spin-down  states along the N{é}el vector $\mathbf{n}$ as the quantization axis, respectively, whereas $\mathrm{P}_{\q}={J}/{\sqrt{\hbar^2v_{0}^2 q^2+J^2}}$ parameterizes the overlap of electron wavefunctions of the two sublattices \cite{PhysRevB.86.245118,PhysRevB.95.134424,PhysRevLett.121.097204}.

In turn, the linearized eigenvalues and the corresponding eigenvectors in the case of  canted AFM are given by,
\begin{align}
&\varepsilon_{\eta,\nu}(\q)=\eta\sqrt{J^2+\hbar^2v^{2}_{0} q^2+2\nu  J m \hbar v_{0} q },\label{linearizedE}\\
&\ket{\Psi_{\eta}^{\nu}}=c_{\eta}^{\nu}\Big(\ket{\psi_{\eta}^{\uparrow}}+i\nu\eta\ket{\psi_{\eta}^{\downarrow}}\Big),
\label{eigenfunction_m}
\end{align}
where
\begin{align}
c_{\eta}^{\nu} = & \dfrac{\sqrt{1+\eta \cos\theta \mathrm{P}_{\q,\nu}}}{2}\nonumber \\
&\times \Bigg\{\sqrt{1+\eta  \mathrm{P}_{\q}}+\eta\nu \dfrac{\varepsilon_{\eta,\nu}-\CMcal{J}m}{\CMcal{J}m+\nu \hbar v_{0} q}\sqrt{1-\eta  \mathrm{P}_{\q}}\Bigg\},
\label{c_coefficient}
\end{align}
with $\mathrm{P}_{\q,\nu}={J}/{\sqrt{J^2+\hbar^2v_{0}^2q^2+2\nu J \hbar v_{0}qm }}$. Equation (\ref{eigenfunction_m}), shows that the eigenvectors in canted AFM case can be written as a linear combination of the eigenvectors in the collinear AFM limit, Eq. (\ref{eigenfunction}).  

In collinear AFM case, the conduction and valence bands are spin degenerate, Fig. \ref{fig:espectrum_AFM}(a).  The net magnetization in canted AFM system leads to the splitting of conduction and valence bands into two subbands with opposite spin-helicity, with the corresponding dispersion curves crossing each other at the Dirac points. This is illustrated in Fig. \ref{fig:espectrum_AFM}(b).  Note, the net magnetization plays here a role similar to the spin-orbit coupling \cite{PhysRevB.80.035438, PhysRevB.101.115412}. The band dispersion around the Dirac point for nonzero magnetization are shown in Fig \ref{fig:espectrum_AFM}(c). When focusing on low energy states (near the Dirac points) one can distinguish two different regimes, depending on the ratio
of the Fermi energy $\varepsilon_{\rm F}$ and the strength of the exchange interaction $J$. In the two-band regime, $\varepsilon_{\rm F} \geq J$, Fermi level intersect two sub-bands with opposite spin-helicity, while in the single-band regime,  $J\sqrt{1-m^2}<\varepsilon_{\rm F} < J$, the Fermi level intersect only one of the subbands.

%------------------------------------------------------------------------------------------------
\section{Emergent magnetic field of skyrmions in a canted AFM system}
\label{sec:em_field}
%------------------------------------------------------------------------------------------------

In noncentrosymmetric magnetic systems, strong DMIs may lead to a helical spin configuration, with the order parameter slowly varying in space, e.g., to formation of skyrmion textures~\cite{PBak_1980,NAKANISHI1980995,BLebech_1989,PhysRevB.52.4389, PhysRevB.88.184422}.
 Accordingly, we assume the N{\'e}el vector is uniform, $\bf n=\bf n_0$, outside the skyrmion and is spatially dependent, $\bf n=\bf n_{r}$,  within the skyrmion region, where it varies according to the skyrmion profile. 
 Without loss of generality, we choose the following profile for describing the corresponding AFM skyrmion in spherical coordinates \cite{PhysRevB.92.014418, USOV1993L290, PhysRevLett.121.097204},
\begin{equation}
\label{eq:n}
\mathbf {n_r} = \left(\cos{\Phi} \sin{\Theta}, \sin{\Phi} \sin{\Theta}, \cos{\Theta} \right),
\end{equation}
where the corresponding polar and azimuthal angles are defined by the following equations: \cite{PhysRevB.95.064426}
\begin{subequations}\label{profile}
\begin{align}
\Theta (r)  &= 2\pi - 4\, \arctan \Bigl(\exp(\frac{4r}{r_{\rm sk}})\Bigr), \\
\Phi &= p \rm{Arg}(x + i y) + c \frac{\pi}{2}.
\end{align}
\end{subequations}
Here, $r_{\rm sk}$ is the skyrmion radius, $r$ is distance from the skyrmion center,  while $p = \pm 1$ and $c = \pm1$ describe skyrmion vorticity and chirality, respectively.

Because of the spatial variation of the N{\'e}el vector inside the skyrmion region, the sd exchange term in Hamiltonian, Eq. (\ref{main_Hamiltonian_b}), is not diagonal anymore. However, the sd exchange term can be diagonalized by performing an appropriate unitary transformation $U(\mathbf{r})$~\cite{TATARA2008213, PhysRevB.77.134407},
\begin{align*}
U^{\dagger}(\mathbf{r})\cdot \Bigg(J \Big[\hat{\tau}_z \otimes \mathbf{n}+\mathbb{1}\otimes \mathbf{m} \Big]\cdot \bsig\Bigg)\cdot U(\mathbf{r})=J\tau_z\otimes\sigma_z.
\end{align*}
The spin rotation operator $U$, that diagonalizes the sd exchange term, is a $4 \times 4$ matrix, which in our case takes the following form:
\begin{equation}
U(\mathbf{r})=\begin{bmatrix}
\tilde{\mathbf{m}}_a(\mathbf{r})\cdot\s  & 0 \\[5pt]
0 & \tilde{\mathbf{m}}_b(\mathbf{r})\cdot\s,
\end{bmatrix},
\end{equation}
where we introduce,
\begin{align*}
&\tilde{\mathbf{m}}_a=\big(\sin(\dfrac{\Theta+\theta}{2})\cos\Phi,\sin(\dfrac{\Theta+\theta}{2})\sin\Phi,\cos(\frac{\Theta+\theta}{2})\big),\\
&\tilde{\mathbf{m}}_b=\big(\sin(\dfrac{\Theta-\theta}{2})\cos\Phi,\sin(\dfrac{\Theta-\theta}{2})\sin\Phi,\cos(\frac{\Theta-\theta}{2})\big).
\end{align*}
With this gauge transformation, the itinerant electrons interacting with the localized spins (nonuniformly polarized) of skyrmions become transformed into electrons that are uniformly spin-polarized. They then interact with an SU(2) gauge field, $\CMcal{A} = i \frac{\hbar}{e} U^\dagger \nabla U$, which is localized around the skyrmion. This gauge field serves as an emerging vector potential, giving rise to the following spin- and sublattice-dependent emergent magnetic field \cite{G.E.Volovik_1987, PhysRevLett.83.3737, PAPANICOLAOU1991425, PhysRevLett.98.246601, PhysRevB.77.134407, PhysRevLett.102.086601},
\begin{eqnarray}
\bv_{\mathrm{em}}=\nabla\times { \CMcal{A}}=
-\dfrac{\hbar}{e}(\nabla \Theta\times\nabla\Phi)\hspace{1.3cm}\nonumber \\
\times
\begin{bmatrix}
\sin(\dfrac{\Theta + \theta}{2}) \s\cdot\tilde{\mathbf{m}}_a & 0 \\[6pt]
0 & \sin(\dfrac{\Theta -\theta}{2}) \s\cdot\tilde{\mathbf{m}}_b 
\end{bmatrix}.
\end{eqnarray}

For our purpose, it is more convenient to define the effective emergent magnetic field acting on  electrons in the spin subband $\nu$ of the band $\eta$;
\begin{equation}
\CMcal{B}_{\mathrm{em},\eta}^{\nu}=\nu\Big(\bv_{\mathrm{em},\eta}^{\nu,\uparrow}-\bv_{\mathrm{em},\eta}^{\nu,\downarrow}\Big),
\label{effective_field_m1}
\end{equation}
where
\begin{equation}
\begin{aligned}
& \bv_{\mathrm{em},\eta}^{\nu,\uparrow}=\vert c^{\nu}_{\eta}\vert^2\bra{\psi^{\uparrow}_\eta} \bv_{\mathrm{em}} \ket{\psi^{\uparrow}_\eta}, \\
&
\bv_{\mathrm{em},\eta}^{\nu,\downarrow}=\vert c^{\nu}_{\eta} \vert^2\bra{\psi^{\downarrow}_\eta} \bv_{\mathrm{em}} \ket{\psi^{\downarrow}_\eta}.
\end{aligned}
\label{eq:b_up_dow_nu}
\end{equation}
After some straightforward calculations, we find,
\begin{equation}
\CMcal{B}_{\mathrm{em},\eta}^{\nu}=\nu |c^{\nu}_{\eta}|^2\cos\theta\; \CMcal{B}_{\mathrm{em}}
\label{effective_field_m2},
\end{equation}
with
\begin{equation}
\CMcal{B}_{\mathrm{em}}=-\dfrac{\hbar}{e}(\nabla \Theta\times\nabla\Phi)\sin\Theta.
\label{effective_field2}
\end{equation}
Without loss of generality, we assume that the Fermi level is in the conduction band. Thus, throughout the remainder of this article, we set $\eta=+1$ and omit this subscript.

%----------------------------------------------------------------------------------------
\section{Semiclassical Boltzmann approach}
\label{sec:Boltzmann}
%----------------------------------------------------------------------------------------
To compute spin accumulation, TCHE, and TSHE due  to skyrmions in canted AFM metals, we utilize the semiclassical Boltzmann formalism in the steady-state regime, where the corresponding Boltzmann 
equation can be written in the form~\cite{Ashcroft};  
\begin{equation}
\begin{aligned}
& \mathbf{v}_{\nu} \cdot \frac{\partial f_{\nu}}{\partial \mathbf{r}} - e \Bigl(\mathbf{E} + \mathbf{v}_{\nu} \times \CMcal{B}_{\mathrm{em}}^{\nu} \Bigr) \cdot \frac{\partial f_{\nu}}{\hbar \partial \mathbf{q}} = \\[5pt]
&
\hspace{2cm}
- \frac{f_{\nu} - \langle f_{\nu}\rangle}{\tau_{\nu}} - \frac{\langle f_{\nu}\rangle - \langle f_{-\nu}\rangle}{\tau_{\rm sf}}.
\end{aligned}
\label{eq:Boltzmann}
\end{equation}
Here, $f_{\nu} = f_{\nu}(\mathbf{r}, \mathbf{q})$ is the nonequilibrium Fermi-Dirac distribution function for electrons with velocity $\mathbf{v}_{\nu}$ in the spin subband $\nu$, while $\ev=E_x \hat{x}$ is the applied electric field. Furthermore,  $\langle ... \rangle$ denotes the angular average over the momentum space, i.e., $\langle f_{\nu} \rangle = \int d^{2}\Omega_{\mathbf{q}} f_{\nu} / \int d^{2}\Omega_{\mathbf{q}}$, where $\Omega_{\mathbf{q}}$ is the solid angle in the momentum space. 
The first term on the right-hand side describes the spin-conserving scattering processes, with $\tau_{\nu}$ being the corresponding spin-dependent relaxation time. In turn, the second term takes into account spin-flip scattering processes, with $\tau_{\rm sf}$ denoting the corresponding spin-flip relaxation time.

Within the linear response theory, the nonequilibrium distribution function can be decomposed into an equilibrium part, $f_{\nu}^{0}$, and a perturbation induced by the electric field and effective emergent magnetic field,
\begin{equation}
\label{eq:fnu}
f_{\nu} = f_{\nu}^{0} - \frac{\partial f_{\nu}^{0}}{\partial \varepsilon} \Bigl(- e \mu_{\nu}(\mathbf{r}) + g_{\nu}(\mathbf{r}, \mathbf{q}) \Bigr),
\end{equation}
where 
$- e \mu_{\nu}(\mathbf{r})$ and $g_{\nu}(\mathbf{r}, \mathbf{q})$ are the isotropic and anisotropic parts of the distribution function, respectively, and $\int d^{2}\mathbf{q}\, g_{\nu}(\mathbf{r}, \mathbf{q}) = 0$.
Inserting Eq. (\ref{eq:fnu}) into the Boltzmann equation (\ref{eq:Boltzmann}), we find the following equations for the odd and even velocity moments of the distribution function \cite{PhysRevB.97.134401}:
\begin{equation}
-e\big(\mathbf{E}-\nabla_{\rv}\mu_{\nu}(\mathbf{r})\big)\cdot \v_{\nu}+\frac{e}{\hbar}\big(\v_{\nu}\times \CMcal{B}_{\rm em}^{\nu}\big)\cdot \dfrac{\partial g_{\nu}(\rv,\q)}{\partial \q}=\dfrac{g_\nu(\rv,\q)}{\tau_\nu},
\label{eq:g1}
\end{equation}
\begin{equation}
\v_{\nu}\cdot \dfrac{\partial g_{\nu}(\rv,\q)}{\partial \rv}=\frac{e}{\tau_{\rm sf}}\big(\mu_\nu-\mu_{-\nu}\big).
\label{eq:g2}
\end{equation}
From Eq. (\ref{eq:g1}), we find
\begin{equation}
\begin{aligned}
&
g_\nu(\rv,\q)=-e\tau_{\nu}\big(\mathbf{E}-\nabla_{\rv}\mu_{\nu}(\mathbf{r})\big)\cdot \v_{\nu} \\
&
\hspace{2cm}-\frac{(e\tau_\nu)^2}{\hbar}\big(\v_{\nu}\times \CMcal{B}_{\rm em}^{\nu}\big)\cdot \dfrac{\partial }{\partial \q}\big(\mathbf{E}\cdot\v_{\nu}).
\end{aligned}
\label{eq:g3}
\end{equation}
Upon inserting Eq. (\ref{eq:g3}) into Eq. (\ref{eq:g2}) we get,
\begin{eqnarray}
\sum_{i,j}\dfrac{\partial^2 \mu_\nu}{\partial x_i\partial x_j}v_{\nu,i} v_{\nu,j}-\dfrac{(e \tau_\nu v_{\nu})^2 }{\hbar q^2} \q\cdot\nabla\Big[(\v_\nu\times  \CMcal{B}^\nu_{\rm em})\cdot\ev\Big]  \nonumber \\
=\frac{\mu_{\nu}-\mu_{-\nu}}{\tau_\nu\tau_{\rm sf}},\hspace{2cm}
\label{eq:mu_diffe}
\end{eqnarray}
where $ v_{\nu,i}$represents the $i^{\rm th}$ component of the electron velocity. 
 At this point we note that including Berry curvature and the related anomalous velocity leads to nonlinear terms. These terms, however, are  not included here. For more details on the influence of the Berry curvature and anomalous velocity, as well as on the nonlinear terms  see the Appendix. 

Having derived Eq. (\ref{eq:g3}) and Eq. (\ref{eq:mu_diffe}), we can now compute spin accumulation, TSHE, and TCHE in different regimes of the canted AFM system. 
We begin with the case when the two spin subbands are occupied and contribute to current.

%----------------------------------------------------------------------------------------
\subsection{Two-subbband regime: $\varepsilon_{\rm F} \geq J$}
\label{sec:two-band}
%----------------------------------------------------------------------------------------

First, we consider the case when the Fermi energy is larger than the sd exchange interaction, $\varepsilon_{\rm F}>J$, and thus electrons from both spin subbands $\nu$ contribute to transport.
To derive the equation, from which  the spin accumulation can be calculated, we first carry out the angular integration in the momentum space
on both sides of Eq. (\ref{eq:mu_diffe}). From this we find,
\begin{equation}
\nabla^2\mu_\nu-\dfrac{\mu_\nu-\mu_{-\nu}}{l_{\nu}^2}=\frac{ev_{0}\tau_\nu}{\varepsilon_\nu}\left(v_{0}+\dfrac{\nu J m}{\hbar q_{\nu,\rm F}}\right)(\nabla\times{\CMcal{B}_{\mathrm{em}}^{\nu}})\cdot\ev,
\label{eq:mu_diffe_2band}
\end{equation}
where $l_{\nu}^2={v_{\mathrm{F}}^2\tau_\nu\tau_{\rm sf}}/{2}$ ($l_{\nu}$ is the spin-flip diffusion length in the subband $\nu$)  and $v_{\mathrm{F}}={v_{0}}\sqrt{1-(J/\varepsilon_\mathrm{F})^2(1-m^2)}$ is the Fermi velocity. 
From the above formula, one can derive equation for the spin accumulation $\delta\mu=(\mu_{-}-\mu_{+})/2$ \cite{PhysRevB.97.134401,ZAREZAD2024171599},
\begin{equation}
\nabla^2\delta\mu-\dfrac{\delta\mu}{\lambda_{\rm sd}^2}=-\frac{e v_{0} \cos\theta}{2\hbar}\ev\cdot\Big[\nabla\times{(\CMcal{B}_{\mathrm{em}}\hat{z})}\Big]\sum_{+,-}\dfrac{\tau_{\pm}|c^{\pm}|^2}{q_{\pm,\rm F}},
\label{eq:diffusion_eq_m1}
\end{equation}
where $\lambda_{\rm sd}$ is the spin-averaged diffusion length, $2/\lambda_{\rm sd}^2=\big(1/l_{+,\rm F}^2+1/l_{-, \rm F}^2\big)$, and $q_{\nu,\mathrm{F}}=\big(-\nu m J+\varepsilon_{\mathrm{F}}\sqrt{1-(J/\varepsilon_{\mathrm{F}})^2(1-m^2)}\big)/(\hbar v_{0})$ is the Fermi wavevector for the spin subband $\nu$.
Introducing the parameters  $\tau=(\tau_{+}+\tau_{-})/2$ and $p_{\tau}=(\tau_{-}-\tau_{+})/(\tau_{+}+\tau_{-})$ for the spin-averaged momentum relaxation time and spin asymmetry of the relaxation time, respectively, Eq.~(\ref{eq:diffusion_eq_m1}) can be rewritten as 
\begin{equation}
\nabla^2\delta\mu-\dfrac{\delta\mu}{\lambda_{\rm sd}^2}=-\dfrac{e\tau E_x }{2\hbar}\dfrac{ v_{0}\cos\theta}{M(q_{+,\rm F},q_{-,\rm F})}\dfrac{d\CMcal{B}_{\rm em}^z}{dy},
\label{eq:diffusion_eq_m2}
\end{equation}
where we defined  the parameter $ M(q_{+,\rm F},q_{-,\rm F}) \equiv M$ as follows:
\begin{align*}
\dfrac{1}{ M}=(1-p_{\tau})\dfrac{|c^+(q_{+,\rm F})|^2}{q_{+,\rm F}}+(1+p_{\tau})\dfrac{|c^-(q_{-,\rm F})|^2}{q_{-,\rm F}}.
\end{align*}

%----------------------------------------------------------------------------------------
\subsubsection{\bf Spin and charge current densities}
%----------------------------------------------------------------------------------------

Having nonequilibrium Fermi distribution, Eq. (\ref{eq:fnu}), one can compute the spin($\nu$)-dependent current density,
\begin{align}
\j_{\nu}&=-\dfrac{e}{(2\pi)^2}\int d^2\q  f_{\nu}(\rv,\q)\v_\nu, \nonumber\\
&=\sigma_\nu \Big(\ev-\nabla_{\rv}\mu_{\nu}-\dfrac{e\tau_{\nu}}{\hbar}\dfrac{v_{0}}{q_{\nu,\mathrm{F}}}\ev\times\CMcal{B}_{\mathrm{em}}^{\nu}\Big),
\end{align}
where $\sigma_\nu=({e^2}/{2h})({\tau_\nu v_{0}q_{\nu,\mathrm{F}}})$ is the charge conductivity of the subband $\nu$. 
Then, the total charge $\j^{\rm ch}$ and spin $\j^{\rm sp}$ current densities can be calculated as $\j^{\rm ch}=\j_{+}+\j_{-}$ and $\j^{\rm sp}=\j_{-}-\j_{+}$, respectively. Finally, the total transverse (Hall) charge and spin current densities are given by the formulas, 
\begin{widetext}
\begin{subequations}\label{current1}
\begin{align}
& \overline{j}^{\rm ch}_y(y)=-\sigma\big(\dfrac{d \overline{\mu}}{dy}+p_{\sigma}\dfrac{d \overline{\delta\mu}}{dy}\big)+\dfrac{e v_{0}\tau\sigma \cos\theta}{2\hbar}
\Big[(1-p_{\tau})(1-p_{\sigma})\dfrac{|c^+|^2}{q_{+}}-(1+p_{\tau})(1+p_{\sigma})\dfrac{|c^-|^2}{q_{-}}\Big]\overline{\CMcal{B}}_{\mathrm{em}}^z E_x,\label{eq:j_ch_m}\\[8pt]
& 
\overline{j}^{\rm sp}_y(y)=-\sigma\big(p_{\sigma}\dfrac{d \overline{\mu}}{dy}+\dfrac{d \overline{\delta\mu}}{dy}\big)+\dfrac{e v_{0}\tau\sigma \cos\theta}{2\hbar}
\Big[(1-p_{\tau})(1-p_{\sigma})\dfrac{|c^+|^2}{q_{+}}+(1+p_{\tau})(1+p_{\sigma})\dfrac{|c^-|^2}{q_{-}}\Big]\overline{\CMcal{B}}_{\mathrm{em}}^z E_x, \label{eq:j_sp_m}
\end{align}
\end{subequations}
%\end{widetext}
where $\sigma=\sigma_{+}+\sigma_{-}$ and $\mu=(\mu_{+}+\mu_{-})/2$ are the total charge conductivity and spin-averaged chemical potential, respectively; and $p_{\sigma}\equiv({\sigma_{-}-\sigma_{+}})/({\sigma_{+}+\sigma_{-}})$ is the spin asymmetry of the conductivity. 
For any function $F(x,y)$, we have defined $\overline{F}(y)=({2L})^{-1}\bigintsss_{-L}^{L} dx F(x,y)$, where $2L$ is the length of the system.

Total spin accumulation at the boundaries can be determined by integrating Eq. (\ref{eq:diffusion_eq_m2}) along the $x$ direction, which gives
\begin{equation}
\dfrac{d^2\overline{\delta\mu}}{dy^2}-\dfrac{\overline{\delta\mu}}{\lambda_{\rm sd}^2}=-\dfrac{e\tau E_x v_{0}\cos\theta}{2\hbar M}\dfrac{d\CMcal{B}_{\rm em}^z}{dy}.
\label{eq:diffusion_eq_m3}
\end{equation}
To solve the differential equations (\ref{current1}) and (\ref{eq:diffusion_eq_m3}), we need to employ the appropriate boundary conditions. Assuming that the AFM nanoribbon has a finite width with open boundary conditions, the transverse component of the charge current density must be zero everywhere, i.e., $\overline{j}^{\rm ch}_{y}(y) = 0$. Imposing this condition on Eq. (\ref{eq:j_ch_m}), we find skyrmion-induced transverse electric field as follows,
\begin{equation}
E_y=-\dfrac{d\overline{\mu}}{dy}=p_{\sigma}\dfrac{d\overline{\delta\mu}}{dy}-\dfrac{e  \tau }{2\hbar}\dfrac{v_{0}\cos\theta}{\Gamma(q_{+, \rm F},q_{-, \rm F})}\overline{\CMcal{B}}_{\mathrm{em}}^z E_x,
\label{eq:open_boundary_condition}
\end{equation}
where we defined $\Gamma (q_{+,\rm F},q_{-,\rm F}) \equiv \Gamma$ as 
\begin{equation}
\dfrac{1}{\Gamma}= 
(1-p_{\tau})(1-p_{\sigma})\dfrac{|c^+(q_{+,\rm F})|^2}{q_{+,\rm F}}-(1+p_{\tau})(1+p_{\sigma})\dfrac{|c^-(q_{-,\rm F})|^2}{q_{-,\rm F}}.\nonumber
\end{equation}

On the other hand, the spin current density must be zero only at edges of the AFM nanoribbon, i.e., $\overline{j}^{\rm sp}_{y}(\pm w)=0$, where $2w$ is the nanoribbon width. Using this condition, we find the general solution of Eq.(\ref{eq:diffusion_eq_m3}) in the form:
%\begin{widetext}
\begin{equation}
\overline{\delta\mu}(y)=-\dfrac{e\tau  v_{0}\cos\theta}{4\hbar M}\Bigg(\dfrac{\sinh(\frac{y}{\lambda_{\rm sd}})\exp(\dfrac{-w}{\lambda_{\rm sd}})}{\cosh(\dfrac{w}{\lambda_{\rm sd}})}\bigintssss_{-w}^{+w}\overline{\CMcal{B}}_{\mathrm{em}}^z\exp(\dfrac{\tilde{y}}{\lambda_{\rm sd}})d\tilde{y}+\bigintssss_{-w}^{+w}\overline{\CMcal{B}}_{\mathrm{em}}^z\dfrac{y-\tilde{y}}{|y-\tilde{y}|}\exp(\dfrac{-|y-\tilde{y}|}{\lambda_{\rm sd}})d\tilde{y}\Bigg)E_x.
\label{eq:delta_mu_m}
\end{equation}
%\end{widetext}
The first integral on the right-hand side of this equation corresponds to the homogeneous solution of the differential equation (\ref{eq:diffusion_eq_m3}), while the second integral denotes its particular solution.
Inserting the expression for spin accumulation, Eq. (\ref{eq:delta_mu_m}), into Eq.(\ref{eq:j_sp_m}), we find the TSH current density,
%\begin{widetext}
\begin{equation}
     \overline{j}^{\rm sp}_y(y)=\sigma (1+p_{\sigma}^2)\dfrac{e\tau  v_{0}\cos\theta}{2\hbar M}\Bigg[
    \dfrac{1}{2\lambda_{\rm sd}}\Bigg(\dfrac{\cosh(\frac{y}{\lambda_{\rm sd}})\exp(\dfrac{-w}{\lambda_{\rm sd}})}{\cosh(\dfrac{w}{\lambda_{\rm sd}})}\bigintssss_{-w}^{+w}\overline{\CMcal{B}}_{\mathrm{em}}^z\exp(\dfrac{\tilde{y}}{\lambda_{\rm sd}})d\tilde{y}
    -\bigintssss_{-w}^{+w}\overline{\CMcal{B}}_{\rm em}\exp(-\dfrac{|y-\tilde{y}|}{\lambda_{\rm sd}})d\tilde{y}\Bigg)-\overline{\CMcal{B}}_{\mathrm{em}}^z\Bigg]E_x,
\label{eq:j_sp_m1}
\end{equation}
\end{widetext}
Note that for $\theta=0$ ($m=0)$ in Eqs. (\ref{eq:open_boundary_condition}), (\ref{eq:delta_mu_m}), and (\ref{eq:j_sp_m1}), the results reduce to those for the collinear AFM case \cite{ZAREZAD2024171599}.

 For numerical calculations one needs to take appropriate values of the asymmetry parameters $p_{\sigma}$, $p_{\tau}$, and spin diffusion length $\lambda_{\mathrm{sd}}$ -- all being generally dependent on the Fermi energy. According to the definition of $\sigma_\nu$ given below Eq. (31), one can write
%\begin{equation}
$p_{\sigma}=(\tau_-q_{-,\mathrm{F}}-\tau_+q_{+,\mathrm{F}})/(\tau_-q_{-,\mathrm{F}}+\tau_+q_{+,\mathrm{F}})$,
%\end{equation}
where the Fermi wavevectors are $q_{\nu,\mathrm{F}}=\big(-\nu m J+\varepsilon_{\mathrm{F}}\sqrt{1-(J/\varepsilon_{\mathrm{F}})^2(1-m^2)}\big)/(\hbar v_{0})$. 
As $\tau_{+}=\tau (1-p_{\tau})$ and $\tau_{-}=\tau (1+p_{\tau})$, the formula for $p_\sigma$ can be written explicitly as 
\begin{equation}
p_{\sigma}=
\dfrac{mJ+p_{\tau}\sqrt{J^2(m^2-1)+\varepsilon_{\mathrm{F}}^2}}{\sqrt{J^2(m^2-1)+\varepsilon_{\mathrm{F}}^2}+p_{\tau}Jm}
\label{p_sigma}.
\end{equation}
In turn, for the parameter $p_\tau$ we take 
$p_\tau=(q_{-,\mathrm{F}}-q_{+,\mathrm{F}})/(q_{+,\mathrm{F}}+q_{-,\mathrm{F}})$, 
which on taking into account the forms of Fermi wavevectors can be written as
\begin{equation}
p_\tau = \dfrac{mJ}{\sqrt{\varepsilon_{\mathrm{F}}^2 -J^2(1-m^2)}}.
\label{p_tau}
\end{equation}
Finally, the spin diffusion length can be then written in the form
 \begin{equation}
 \lambda_{\mathrm{sd}}=\lambda_0 \sqrt{1-J^2/\varepsilon_{\mathrm{F}}^2(1-m^2)}\sqrt{1-p_\tau^2},
 \label{sd_length}
 \end{equation}
 where $\lambda_0=v_0\sqrt{\tau_{\mathrm{sf}}\tau/2}$.
 We note that the description based on the  Boltzmann equation and relaxation time approximation (including spin diffusion length), is not accurate  in the vicinity of the Fermi energy $E_{\mathrm{F}} =  J$, where Fermi contour of the band marked as $\nu =+ $ shrinks to a Dirac point (k=0) in the 2D Brillouin zone. 
 
\begin{figure}[t]
  \includegraphics[width=.38\textwidth]{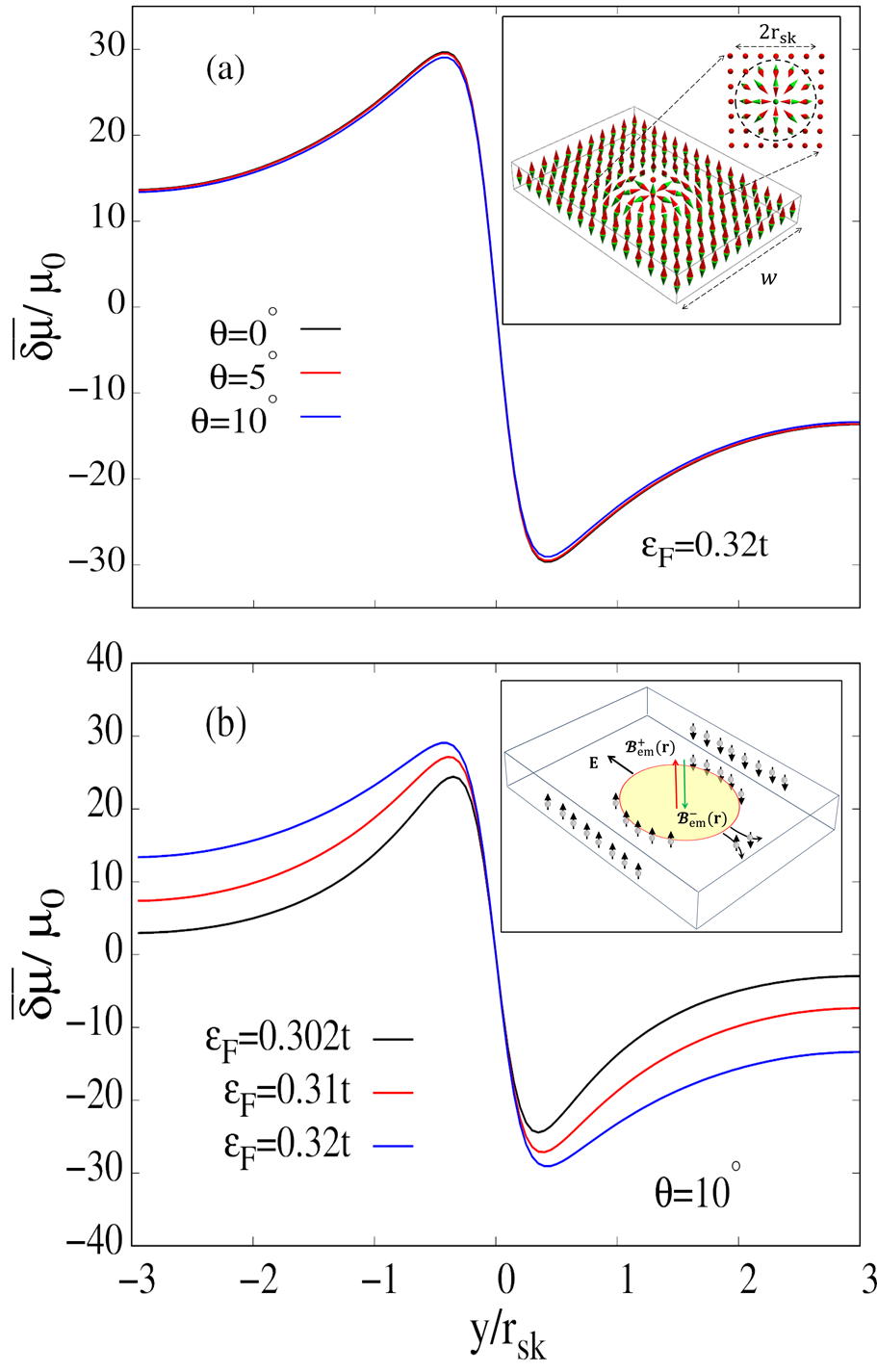}
  \caption{
  Spatial variation of the spin accumulation induced by a single AFM skyrmion for different values of the canting angle $\theta$ (a); and of the Fermi energy $\varepsilon_{\rm F}$ (b). Here, $\mu_0=(r_{\rm sk} e \tau v_{0}^2B_0 E_x )/(16 t)$, $ w=3r_{\rm sk}$, $\lambda_{0}=5r_{\rm sk}$, and $J=0.3t$.
  The inset in (a) shows schematically the stripe of width of $w$, with a single AFM skyrmion of a radius $r_{\mathrm{sk}}$. 
   In turn, the inset in (b) shows schematic of the device for transport measurements: by applying an external electric field, electrons move through a nontrivial magnetic texture and experience a spin-dependent emergent magnetic field, which creates an effective Lorentz force that changes sign for opposite spins and results in transverse spin accumulation.}
  \label{fig:mu}
\end{figure}

 In Fig.~\ref{fig:mu} we illustrate  the spatial variation of the normalized spin accumulation, Eq. (\ref{eq:delta_mu_m}), for indicated parameters, where we have defined 
$B_0=(\pi r_{\rm sk}^2)^{-1}\Phi_{\mathrm{B}}$ with $\Phi_{\mathrm{B}}=\int_{-r_{\rm sk}}^{r_{\rm sk}}d^2\rv \CMcal{B}_{\rm em}^z(\rv)$. The asymmetry parameter $p_\tau$ is determined according to the formula (\ref{p_tau}). 
The normalized spin accumulation is plotted for 
indicated values of the canting angle $\theta$ in Fig.~\ref{fig:mu}(a),
and for indicated values of the Fermi energy and fixed $\theta$ in Fig.~\ref{fig:mu}(b). The spin accumulation is zero in the center of the skyrmion and reaches its minimum (maximum) value around $y= 0.35 r_{\rm sk}$ ($y= -0.35 r_{\rm sk}$), and then its absolute value decreases and saturates at the edges of the AFM stripe.
As follows from Fig.~\ref{fig:mu}(a),  the spin accumulation shown there is practically independent of the canting angle for $\theta \le 10^{\degree}$ (or equivalently independent of the net magnetization, $m=\sin\theta$).
In turn, the absolute magnitude of  the spin accumulation increases with the increasing Fermi energy, see Fig.~\ref{fig:mu}(b). This behavior is associated with the variation of $p_\tau$ with the Fermi energy.
 
\begin{figure}[t]
  \includegraphics[width=.38\textwidth]{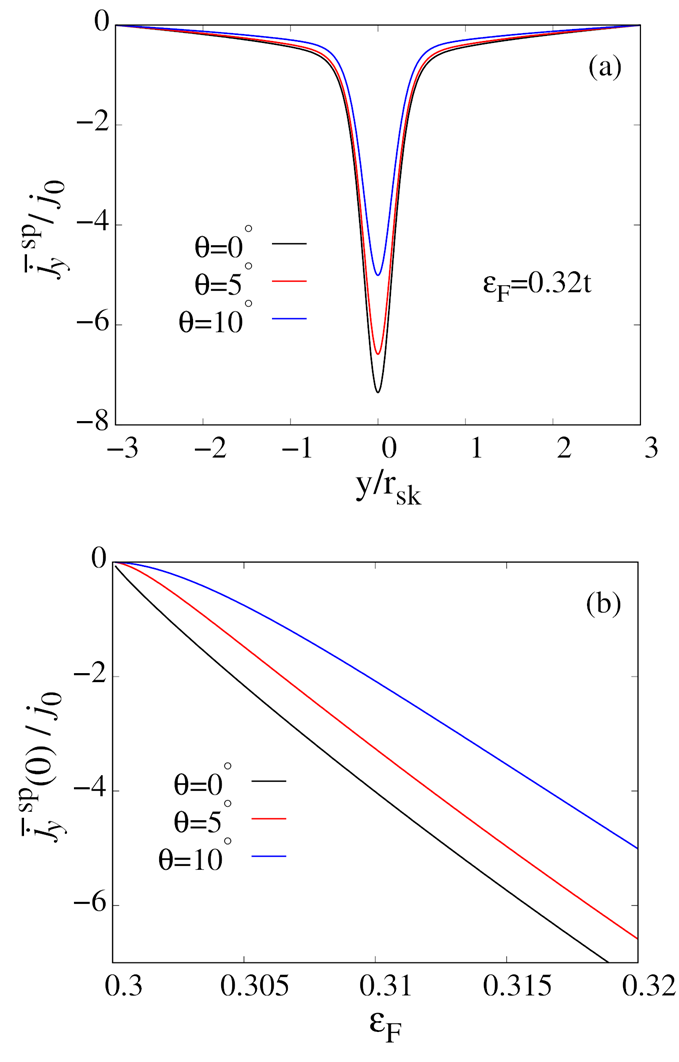}
  \caption{
  Spatial variation of the transverse spin current density generated by a single skyrmion for different values of the canting angle $\theta$ (a). The transverse spin current density in the skyrmion center, $y=0$, as a function of $\varepsilon_{\mathrm{F}}$ (b). Here $j_0= e^3\tau^2 v_{0}^2 E_x B_0/(32\pi\hbar^2)$, $ w=3r_{\rm sk}$, $\lambda_0=5r_{\mathrm{sk}}$ and $J=.3t$}
  \label{fig:j1}
\end{figure}

Figure \ref{fig:j1}(a) illustrates the spatial variation of the transverse spin current density, Eq. (\ref{eq:open_boundary_condition}), for different values of the canting angle. The absolute value of the spin current density reaches a maximum in the skyrmion center, and then decreases with increasing distance from the skyrmion center and monotonically vanishes at the AFM nanoribbon edges. The net magnetization in the system reduces the absolute magnitude of the spin current density (especially around the skyrmion center), as follows fom  Fig.~\ref{fig:j1}(a).  

Figure \ref{fig:j1}(b) shows the spin current density in the skyrmion center as a function of the Fermi energy, $\varepsilon_{\mathrm{F}}$, for the same three canting angles as 
in Fig.~\ref{fig:j1}(a). From this figure follows that the spin current in the skyrmion center vanishes for $\varepsilon_{\mathrm{F}}=J$, independently of the tilting angle. With increasing $\varepsilon_{\mathrm{F}}$, the magnitude of spin current increases monotonously with increasing $\varepsilon_{\mathrm{F}}$. As in Fig.~\ref{fig:j1}(a), the spin current depends on the canting angle and is the largest (negative) one for zero canting angle, i.e., in the strictly collinear antiferromagnet. We  recall that the asymmetry parameters $p_\sigma$ and $p_\tau$, as well as the spin diffusion length, depend on the Fermi energy in Fig.~\ref{fig:j1}(b) and are calculated following Eqs~(\ref{p_sigma}, \ref{p_tau} and \ref{sd_length}).   

%----------------------------------------------------------------------------------------
\subsubsection{\bf Topological charge Hall resistivity}
%----------------------------------------------------------------------------------------

The charge Hall resistivity, generated by skyrmions, is called TCH resistivity and is given by \cite{ZAREZAD2024171599},
\begin{eqnarray}
&\rho_{yx}=\dfrac{\overline{E}_y}{j_x}=-\dfrac{e \tau v_{0} \cos\theta}{8 L w\hbar\sigma} \dfrac{\Phi_{\mathrm{B}}}{\Gamma (q_{+, \rm F},q_{-, \rm F})} \nonumber \\
&\times \Big[1+p_{\sigma}\dfrac{\Gamma(q_{+, \rm F},q_{-, \rm F})}{M (q_{+, \rm F},q_{-, \rm F})}\int d^2\rv\dfrac{\CMcal{B}_{\rm em}(\rv)\cosh(\frac{y}{\lambda_{\rm sd}})}{\Phi_{\mathrm{B}} \cosh(\frac{w}{\lambda_{\rm sd}})}\Big].\hspace{0.3cm}
\label{eq:rho_m1}
\end{eqnarray}
In the untilted limit, $\theta=0$, this expression reduces to the expression for a collinear AFM system \cite{ZAREZAD2024171599}.

\begin{figure}[t]
  \includegraphics[width=.85\columnwidth]{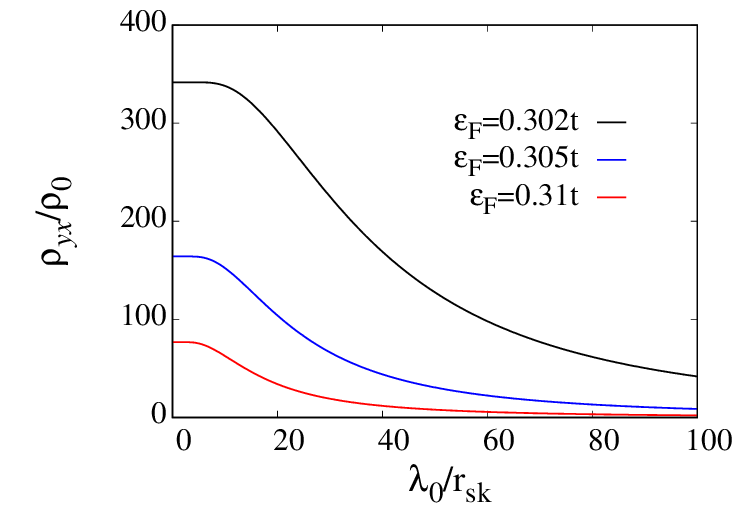}
  \caption{The TCH resistivity in the two-subband regime as a function of $\lambda_0$ and for different values of the  Fermi energy. Here $\rho_0=\pi/(2Lw) (v_{0}^2\hbar^2/t^2)\Phi_{\mathrm{B}}/e$, $J=0.3t$, $\theta=5\degree$, and $w=6r_{\rm sk}$.}
  \label{fig:roh}
\end{figure}

Figure \ref{fig:roh} illustrates the TCH resistivity as a function of the spin-diffusion length $\lambda_{0}$ for indicated values of the Fermi energy $\varepsilon_{\mathrm{F}}$ and for the canting angle  $\theta=5\degree$. This figure shows, that the TCH resistivity decreases with increasing spin diffusion length. Moreover, this figure also shows that the TCH resistivity decreases with increasing Fermi energy, which is associated with decreasing spin asymmetry of the relaxation times with increasing $\varepsilon_{\mathrm{F}}$.
This behaviour is similar to that in the collinear AFM case \cite{ZAREZAD2024171599}.

\begin{figure}[t]
	\centering
\includegraphics[width=.4\textwidth]{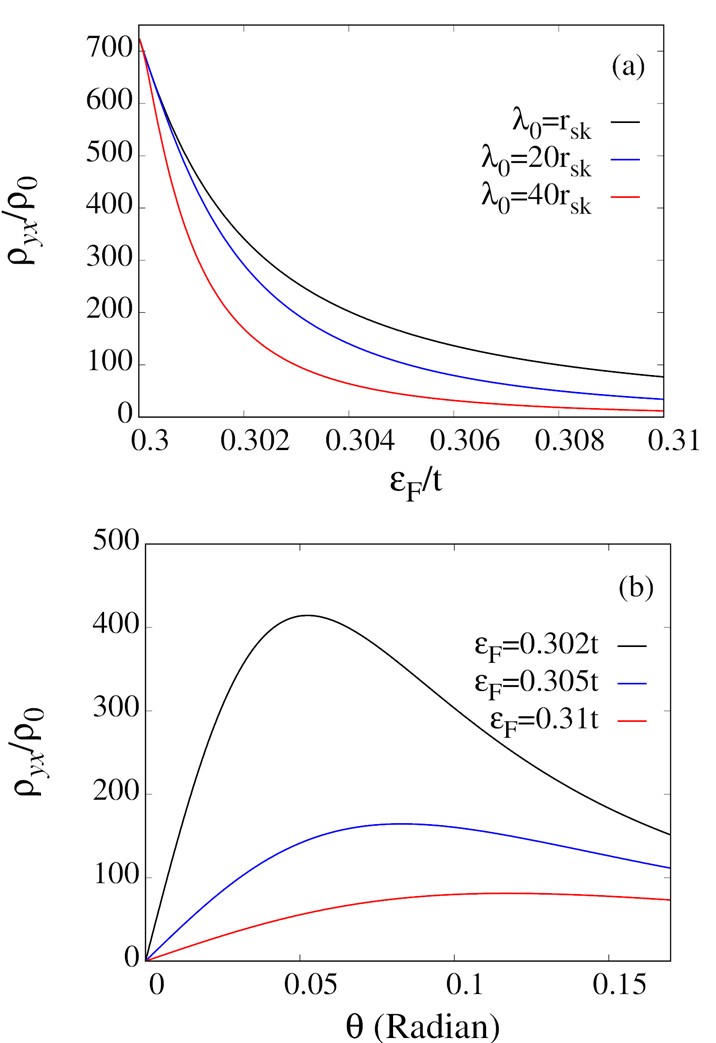}
\caption{TCH resistivity in the two-subband regime for different values of $\lambda_0$ and $\theta=5\degree$, presented as a function of the Fermi energy (a). TCH resistivity as a function of the tilting angle $\theta$ for indicated values of the Fermi energy $\varepsilon_{\rm F}$ and $\lambda_{0}=r_{\rm sk}$ (b). Here, $J=0.3t$, and $w=6r_{\rm sk}$.}   
	\label{fig:fig_roh_m_ef}
 \end{figure}

In Fig. \ref{fig:fig_roh_m_ef}~(a) we explicitly plot the TCH resistivity as a function of the Fermi energy. This figure  shows that increasing the Fermi energy leads to a smooth decrease in the absolute magnitude of the TCH resistivity. This decrease follows from the reduced asymmetry of the two subbands with increasing Fermi energy, which is rather obvious  from the corresponding dispersion curves, see Fig. \ref{fig:espectrum_AFM}.  This behaviour is generally similar to that in the collinear AFM case \cite{ZAREZAD2024171599}.  In turn, from Fig.~\ref{fig:fig_roh_m_ef}(b) follows that the  TCH resistivity vanishes for zero  tilting angle, $\theta =0$, and then grows with increasing $\theta$, reaches a maximum for $\varepsilon_{\rm F}$ slightly above $J$, and then decreases with a further increase in $\theta$. We note, that the absence of TCH effect for $\theta =0$ is due to PT symmetry and absence of additional scattering processes that violate this symmetry. However, we note that even in the collinear AFM case, an asymmetry in relaxation times may lead to a nonzero TCH resistivity, as shown in our earlier work \cite{ZAREZAD2024171599}.

%----------------------------------------------------------------------------------------
\subsection{\bf Single-subband regime: $\varepsilon_{\rm F} < J$}
%----------------------------------------------------------------------------------------

\begin{figure}[t]
	\centering
		\includegraphics[width=.45\textwidth]{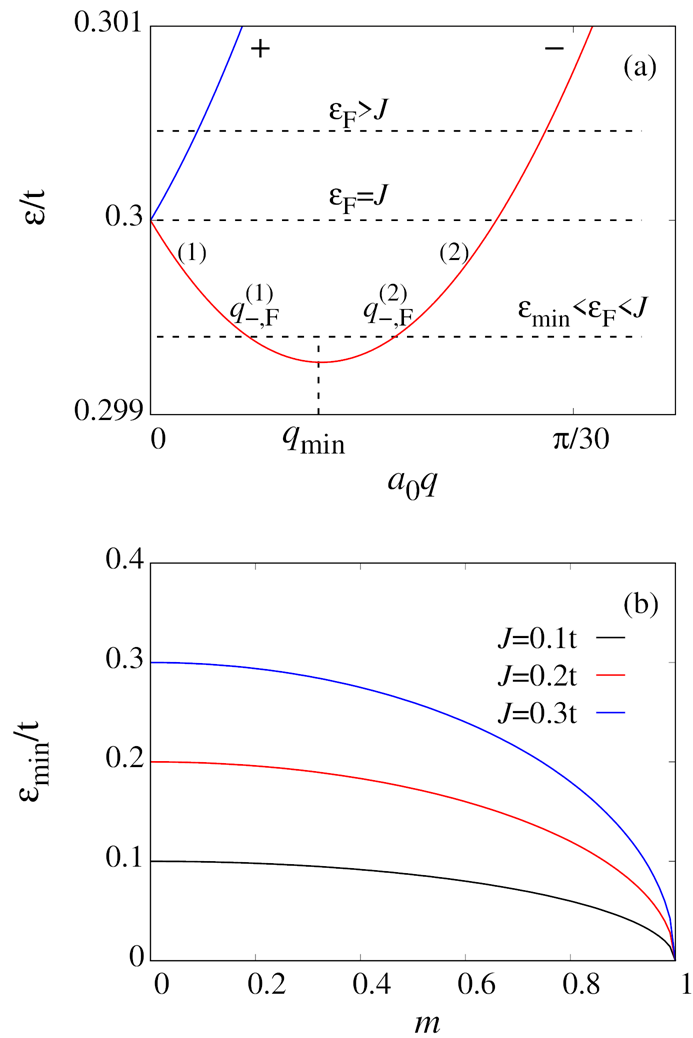}
	\caption{ 
(a): Two conduction subbands around the  $K_+$ point. In the two-subbband regime, 
$\varepsilon_{\rm F}>J$, the Fermi energy intersects the dispersion curve of subband $\nu$ at  $q_{\nu,\mathrm{F}}=(-\nu m J+\varepsilon_{\mathrm{F}}\sqrt{1-(J/\varepsilon_{\mathrm{F}})^2(1-m^2)})/(\hbar v_{0})$. However, when the Fermi energy is smaller than the sd exchange interaction, the Fermi surface intersect only one conduction subband at $ q^{(\alpha )}_{-,\rm F}=\Big(J m+(-1)^{\alpha}\varepsilon_{\rm F}\sqrt{1-(J/\varepsilon_{\rm F})^2(1-m^2)}\Big)/(\hbar v_{0})$ for $\alpha =1,2$. (b): Minimum energy as a function of magnetization for different values of exchange interactions.  }
	\label{fig:espectrum_AFM_single_band}
\end{figure}

When the Fermi energy is smaller than the sd exchange interaction $J$, $\varepsilon_{\mathrm{F}} <J$, only electrons in one spin subband, $\nu=-1$, contribute to the transport, see Fig.~\ref{fig:espectrum_AFM_single_band}(a). 
Accordingly, our main interest is in the TCHE, and therefore we limit our further consideration to the charge transport.  
To calculate charge current we artificially divide the part of dispersion curve in Fig.~\ref{fig:espectrum_AFM_single_band}(a) corresponding to the energy smaller than $J$  into two branches.  
In the branch for $q < q_{\rm min}=Jm/\hbar v_0$, denoted as $\alpha=1$, the energy decreases with increasing $q$, while in the branch for $q>q_{\min}$, marked as $\alpha =2$, the energy increases with  increasing $q$. Here, $q_{\min}$ is the wave number corresponding to the band minimum, $\varepsilon_{\min}=J\sqrt{1-m^2}$. The energy of band minimum is plotted in Fig.~\ref{fig:espectrum_AFM_single_band}(b) as a function of the magnetization $m$. Note, the band gap in the spectrum is closed for $m=1$.
The two Fermi contours correspond to the Fermi wavevectors $q_{-,\mathrm{F}}^{(1)}$ and $q_{-,F}^{(2)}$, as shown in  Fig.~\ref{fig:espectrum_AFM_single_band}(a), with $q_{-,\mathrm{F}}^{(1)} < q_{-,\mathrm{F}}^{(2)}$ and $ q^{(\alpha )}_{-,\rm F}=\Big(J m+(-1)^{\alpha}\varepsilon_{\rm F}\sqrt{1-(J/\varepsilon_{\rm F})^2(1-m^2)}\Big)/(\hbar v_{0})$ for $\alpha =1,2$.   
The Fermi velocity in the contour $\alpha$ for the Fermi level $J\sqrt{1-m^2}<\varepsilon_{\rm F}<J$, is given by,
\begin{equation}
\v_{\alpha,\mathrm{F}}(\varepsilon)={(-1)^{\alpha}v_{0}}\sqrt{1-\big(\frac{J}{\varepsilon_\mathrm{F}}\big)^2(1-m^2)}\hat{\q}
\label{velocity_single_band},
\end{equation}
where $\hat{\q}$ is a unit vector along $\mathbf {q}$. Thus, 
according to the above relation (\ref{velocity_single_band}), the electron group velocities for these two contours are opposite for the same $\hat{\q}$. 

To study TCH resistivity in the single-band regime, we need to formulate the set of relevant transport equations and solve the corresponding Boltzmann equation.
To do this, we treat the two Fermi contours separately, like two states of a \emph{pseudo-spin}.   
Moreover, we distinguish between scattering processes that leave electrons upon scattering in the same Fermi contour $\alpha$ (intra-contour scattering), and scattering processes associated with a change of the Fermi contours, i.e., the inter-contour  scattering.
Treating the branch index as a pseudo-spin index, 
we can map the model on the real spin model used above for the two-subband case. Accordingly, the relaxation time for intra-contour (inter-contour) scattering corresponds to the spin-conserving (spin-mixing) relaxation times in the two spin-subbands model. The chemical potentials in the two contours satisfy an equation similar to that in the model discussed above. Following this similarity, one can write the diffusion equation for the {contour accumulation}  $\delta\mu=(\mu_2-\mu_1)/2$ as,
\begin{equation}
\nabla^2 \overline{\delta\mu}-\dfrac{\overline{\delta\mu}}{\lambda_{\mathrm{mix}}^2}=-\dfrac{e\tau v_{0}\cos\theta }{2\hbar M(q_{\mathrm{F}}^{(1)},q_{\mathrm{F}}^{(2)})}\dfrac{d \overline{\CMcal{B}}_{\mathrm{em,z}}}{dy}E_x,
\label{mu_single_band1}
\end{equation}
where  $\lambda_{\mathrm{mix}}$ is the diffusion length between two inter-contour scatterings, 
and plays the same role as the spin diffusion length in the two spin-subband model. For simplicity, we skipped here (and in the following) the  lower index $\nu =-$ and replaced  $q_{-,\mathrm{F}}^{(1,2)}$ by $q_{\mathrm{F}}^{(1,2)}$. 

Similarly as in the two-band case, we define the energy dependent asymmetry coefficients $p_\sigma$ and $p_\tau$, $p_{\tau}=(\tau_2-\tau_1)/(\tau_1+\tau_2)$ and $p_{\sigma}=(\sigma_2-\sigma_1)/(\sigma_1+\sigma_2)$, as well as the diffusion length $\lambda_{mix}$ (and also the relevant $\lambda_0$).   
Finally, we find the total charge current density $\j^{\rm ch}_y=\j_{1,y}+\j_{2,y}$, in the form
\begin{eqnarray}
\overline{\j}^{\rm ch}_y=-\sigma\Bigg[\dfrac{d \overline{\mu}}{dy}+p_{\sigma}\dfrac{d\overline{\delta\mu}}{dy}\Bigg]+\dfrac{ev_{0}\sigma \tau \cos\theta}{2\hbar}\Bigg[(1-p_\tau)(1-p_\sigma) \nonumber \\
\times \dfrac{|c^-(q_{\rm F}^{(1)})|^2}{q_{\rm F}^{(1)}}-(1+p_\tau)(1+p_\sigma)\dfrac{|c^-(q_{\rm F}^{(2)})|^2}{q_{\rm F}^{(2)}}\Bigg]E_x\overline{\CMcal{B}}_{\mathrm{em}}^z,\hspace{0.75cm}
\end{eqnarray}
where $\sigma=\sigma_1+\sigma_2$,  and  $\sigma_\alpha=(e^2/2h)\tau_\alpha v_\alpha k_{\alpha,\mathrm{F}}$ 
is the charge conductivity in the channel $\alpha$. 

\begin{figure}[t]
	\centering
\includegraphics[width=.4\textwidth]{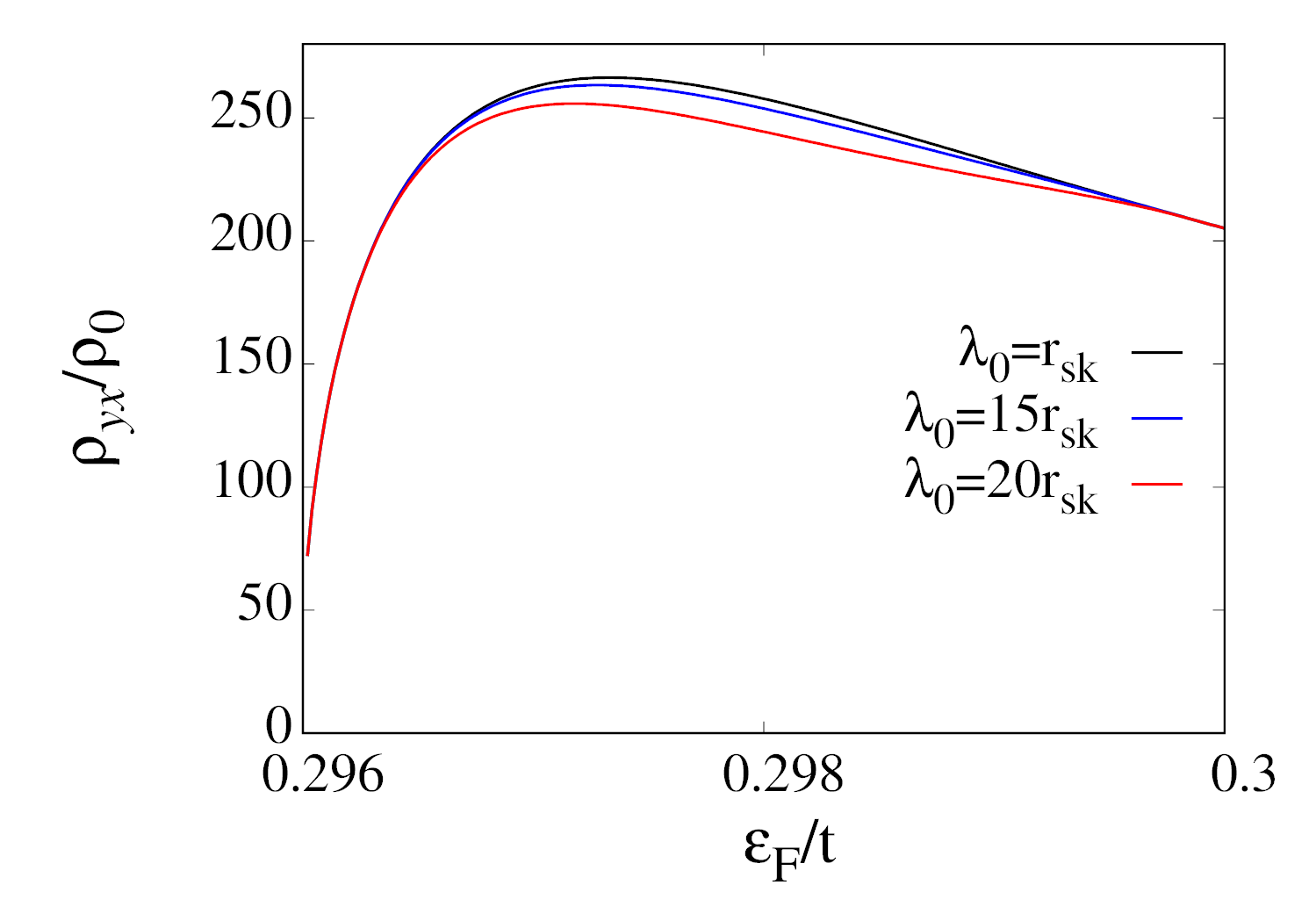}
\caption{The TCH resistivity in the single subband regime shown  as a function of the Fermi energy for different values of the parameter $\lambda_0$, and for $m=0.163$, $J=0.3t$, $w=6r_{\rm sk}$.}   
	\label{fig:fig_roh_ef_single}
 \end{figure}
 
Similarly to the two-subband regime, by imposing the open boundary condition, we can solve the diffusion equation (\ref{mu_single_band1}) and derive  the expression for the  TCH resistivity.
The calculated TCH resistivity is shown in Fig.~\ref{fig:fig_roh_ef_single} as a function  of the Fermi energy for indicated values of the diffusion length $\lambda_0$.
This figure shows that the TCH resistivity increases when the Fermi energy grows from the band minimum, reaches a maximum and then decreases when the Fermi energy approaches the limit $\varepsilon_{\mathrm{F }}=J$.

\subsection{Transition from the two-subband to one-subband regime}

When considering transition from the two-subband to one-subband regimes with reducing the Fermi energy, one needs to pay special attention to the parameters of the model. This follows from the fact that for $E_{\mathrm{F }}=J$, the Fermi energy reaches the bottom of the upper band, and therefore the number of electrons in this band, that participate in transport, goes to zero with $q_{+,{\mathrm{F }}}\to 0$.  In this limit some parameters of the Boltzmann approach are now well defined. To get reliable results, one needs to take energy dependent transport parameters, as described above (see Eqs.(\ref{p_sigma}, \ref{p_tau}, \ref{sd_length}) for $\varepsilon_\mathrm{F}\ge J$.

Similar conditions have to be obeyed when approaching the limit $\varepsilon_{\mathrm{F }} = J$ from the lower energy side $\varepsilon_\mathrm{F}\le J$. However, instead of the two subbands, we have now two wings of the lower energy band, marked  as $\alpha =1$ and $\alpha =2$ in Fig. \ref{fig:espectrum_AFM_single_band}. Behavior of the topological Hall resistivity as a function of the Fermi energy is shown in Fig. 9. 
\begin{figure}[t]
  \includegraphics[width=.85\columnwidth]{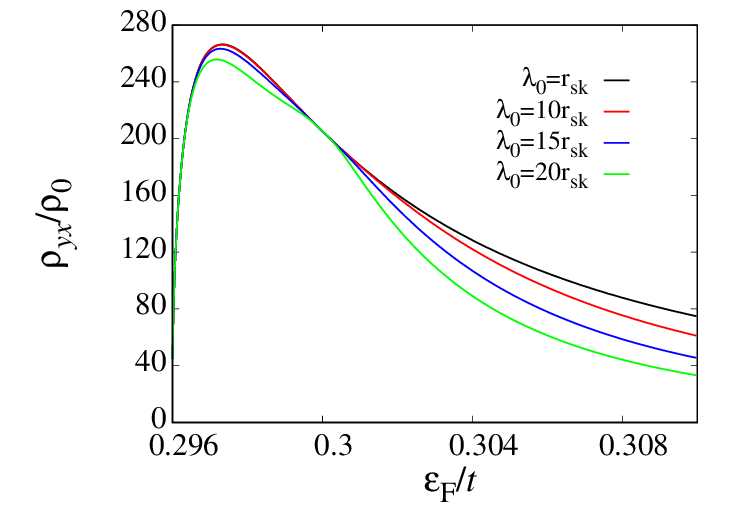}
  \caption{ The TCH resistivity as a
          function of the Fermi energy in canted AFM for different values of the $\lambda_0$ (assumed the same in both regions). Here, $J=0.3t$, $w=6r_{\rm sk}$ , and $m=0.163$.}
  \label{fig:roh-connection}
\end{figure}
This figure indicates that there is a continuous transition between the two regions. Some weak irregularities can be seen around $\varepsilon_{\mathrm{F }}=J$, which follow from the fact that the Boltzmann approach does not properly describes this limit. On the other hand, spin current was studied in the two-band model and it was shown that it vanished for the Fermi energy approaching the value $\varepsilon_{\mathrm{F }} =J$, see Fig.\ref{fig:j1}(b). Since in the one band limit, only one spin orientation is available for transport, the pure spin current is absent.
%-----------------------------
\section{Summary}
\label{sec:Summary}
%-----------------------------

In this paper, we have analyzed the TSHE and TCHE induced by skyrmion in canted AFMs. Such materials are also referred to as weak ferromagnets, due to a small net magnetic moment arising from the canting of the AFM sublattice moments. 
This canting breaks the PT symmetry, not only enhancing the TSHE but also generating a TCHE, which is typically absent in collinear AFM systems with PT symmetry.
The canting of AFM magnetic moments remarkably changes the corresponding electronic structure, and breaks the spin degeneracy of the conduction and valence bands, resembling a Rashba-type spin-orbit splitting of spin-subbands in a two-dimensional electron gas. Employing the semiclassical Boltzmann formalism, we investigated TCHE and TSHE in different energy scales of the system. Note, we have considered only one Dirac point (K) as the contributions from both (K and K') Dirac points are equal.

The obtained  results are consistent with other works on topological spin and charge Hall effects in AFM systems. We mention here the work by Akosa {\it et al}~\cite{PhysRevLett.121.097204}, who analyzed the topological spin and charge Hall effects  within the tight-binding model of  an
AFM square lattice. Using the Landauer–Büttiker formalism, they found a zero topological charge Hall effect and a nonzero topological spin Hall effect in such
a system. 
In turn, Nakazawa {\it et al.}~\cite{nakazawa2023topological} used a similar formalism and showed that the vector chirality formed by the AFM Néel vector leads to a finite topological spin Hall effect in bulk AFM systems with half-skyrmions or merons. 

In our work, we considered a weak ferromagnetic system and took into account intrinsic spin asymmetry of the electric conductivity and relaxation times. We have shown, this asymmetry leads to the topological spin and charge Hall effects in the canted systems, while only spin Hall effect appears then in the collinear AFM.  This is consistent with Ref.~\cite{ PhysRevB.101.174432}, were the topological charge Hall effect was also found in AFM square lattice with canted sublattice magnetic moments. The authors of Ref.~\cite{ PhysRevB.101.174432}, however, included the ferromagnetic moment perturbatively and used a different approach based on the Kubo formalism.

%-----------------------------
\section*{Acknowledgments}
%----------------------------- 
This work has been supported by the Norwegian Financial Mechanism 2014 - 2021 under the Polish - Norwegian Research Project NCN GRIEG “2Dtronics” no. 2019/34/H/ST3/00515.
%%%%%%%%%%%% References %%%%%%%%%%

\appendix \label{appendix}

\section{Contribution of anomalous velocity}
In this appendix, we show that contribution of the anomalous velocity to the topological charge Hall effect is beyond the linear response theory, considered in this work.
Taking into account Berry curvature and anomalous velocity, 
dynamics of electrons is described by the following semiclassical equations \cite{PhysRevB.59.14915};
\begin{align}
&\hbar \dot{\q}_{\nu}=-e (\ev+\dot{\mathbf{r}}_{\nu}\times \CMcal{B}^{\nu}_{\mathrm{em}}),\label{momentum}\\
&\hbar \dot{\mathbf{r}}_{\nu}=\dfrac{\partial \varepsilon_{\nu}}{\partial \q}-\hbar \dot{\q}_{\nu}\times \Omega_{\nu},
\label{position}
\end{align}
where $\Omega_{\nu}^{z}$  is the Berry curvature, defined as the curl of the Berry connection \cite{Vanderbilt_2018}: 
\begin{equation}
\Omega_{\nu}^{z}=\partial_{q_x} \CMcal{A}_{n,y}-\partial_{q_y} \CMcal{A}_{n,x}=-2\Im \left\langle \partial_{q_x}\Psi^{\nu}|\partial_{q_y}\Psi^{\nu}
\right\rangle ,  
\end{equation}
with $\boldsymbol{\CMcal{A}}_{\nu}= \left\langle \Psi^{\nu}|i\nabla_{\q}\Psi^{\nu}\right\rangle $ being  the Berry connection and $\Psi^{\nu}$ denoting the eigenvector in the canted AFM (see Eq.(12) in main text). In our case, 
$\Omega_{\nu,xy}$ is given as;
\begin{equation}
\Omega_{\nu,xy}=\dfrac{1}{2q}\dfrac{\partial }{\partial_q}\Big[\Big((\dfrac{\varepsilon_{\nu}-\CMcal{J}\cos\theta}{\CMcal{J}\sin\theta-\nu|\gamma_{\q}|})^2-1\Big)(1+\cos\theta \mathrm{P}_{q,\nu})\Big]
\end{equation}
where $\mathrm{P}_{q,\nu}=\CMcal{J}/\sqrt{\CMcal{J}^2+|\gamma_q|^2+2\nu\CMcal{J}\vert\gamma_{\q}\vert\sin\theta }$.  Equations (\ref{momentum}) and (\ref{position}) give: 
\begin{equation}
\dot{\mathbf{r}}_{\nu}=\omega_{\nu}\Big(\dfrac{1}{\hbar}\dfrac{\partial \varepsilon_{\nu}}{\partial \q}+\dfrac{e}{\hbar}\ev \times \Omega_{\nu}\Big),\label{velocity}
\end{equation}
with $\omega_{\nu}=\dfrac{1}{1+(e/\hbar)\Omega_{\nu}\cdot \CMcal{B}^{\nu}_{\mathrm{em}}}$.
Using the velocity modified with geometric phase, Eq. (\ref{velocity}), in the Boltzmann equation (Eq.~(22) in the main text), we obtain the following relation for the nonequlibrium distribution function;
\begin{eqnarray}
g_{\nu}=e\tau_{\nu}\omega_{\nu}\Big(\dfrac{1}{\hbar}\dfrac{\partial \varepsilon_{\nu}}{\partial \mathbf{q}}-\dfrac{e}{\hbar}(\ev\times\Omega_{\nu})\Big)\cdot\nabla_{\mathbf{r}} \mu_{\nu} 
-e\tau_{\nu}\dfrac{1}{\hbar}\mathrm{E}\dfrac{\partial \varepsilon_{\nu}}{\partial q_x} \nonumber \\-\dfrac{e^2\tau_{\nu}^2\omega_{\nu}}{\hbar}\Big(\dfrac{1}{\hbar}\dfrac{\partial \varepsilon_{\nu}}{\partial \q}\times\CMcal{B}^{\nu}_{\mathrm{em}}\Big)\cdot \Big(\dfrac{\partial }{\partial \q}(\dfrac{1}{\hbar}\ev\cdot \dfrac{\partial \varepsilon_{\nu}}{\partial \q})\Big).\hspace{1cm}\label{distribution_func} 
\end{eqnarray}
In the presence of the anomalous velocity, the first and third terms in the distribution function are modified by $\omega_{\nu}$. Since the gradient of chemical potential in the first term of the distribution function is induced by the magnetic field, and because of the emergent magnetic field in the third term, by expanding $\omega_{\nu}\approx 1+(e/\hbar)\Omega_{\nu}\cdot \CMcal{B}_{\mathrm{em}}^{\nu}+(e/\hbar)^2(\Omega_{\nu}\cdot \CMcal{B}_{\mathrm{em}}^{\nu})^2+\cdots$, only the first term of the expansion contributes to the distribution function, while the higher-order terms in the expansion of $\omega_{\nu}$ create nonlinear terms that are negligible. Thus, by considering the first term of the expansion, the distribution function reduces to the one in the absence of the anomalous velocity.
Furthermore, the presence of anomalous velocity does not affect the structure of the diffusive equation for spin accumulation. The only effect of the emerging Berry phase on the diffusion equation is the modification of the diffusion length scale, changing from $(1/l_{+,\mathrm{F}}^2+1/l_{-,\mathrm{F}}^2)$ to $\Big(1/(\omega_{+}^3l_{+,\mathrm{F}}^2)+1/(\omega_{-}^3l_{-,\mathrm{F}}^2)\Big)$.

For the transverse current density due to the anomalous velocity, we use following relation; 
\begin{equation}
j_{\nu}^{an}=-\dfrac{e}{(2\pi)^2}\int dq^2(\dfrac{e}{\hbar}\omega_{\nu}\ev \times \Omega_{\nu})(-\dfrac{\partial f_{\nu}^0}{\partial \varepsilon})g_{\nu}
\end{equation}
where $f_{\nu}^0$ is the equilibrium distribution function. Using Eq. (\ref{distribution_func}), we find
\begin{equation}
j_{yx}^{an}=\dfrac{e^2}{h}\dfrac{e^2\mathrm{E}^2}{\hbar^2 v_{\mathrm{F}}}\sum_{\nu} \tau_{\nu}q_{\nu,\mathrm{F}}\omega_{\nu}^2\Omega_{\nu,xy}^2\dfrac{\partial \mu_{\nu}}{\partial y},
\label{j_anomalous}
\end{equation}
where $a_0$ is the lattice constant. As observed from Eq. (\ref{j_anomalous}), the anomalous velocity contributes nonlinearly to the transverse charge current. Since we are investigating only the linear response in this paper, we discard the impact of the anomalous velocity.

Equation (\ref{j_anomalous}) demonstrates that the topological charge Hall current resulting from the anomalous velocity, is nonlinear in electric field and is proportional to the square of the Berry curvature, $\Omega_{\nu}^2$. Consequently, despite the Berry curvatures at the K and K' symmetry points having opposite signs, the total nonlinear charge Hall current, which is the sum of contributions from the two Dirac cones, remains finite. This contribution has been omitted here; however, in some specific situation, where the linear term is negligible, the nonlinear contribution may play an essential role.

\bibliography{Draft-THE_PRB}% Produces the bibliography via BibTeX.

\end{document}